\documentclass[pre,preprintnumbers,amsmath,amssymb,twocolumn, nofootinbib]{revtex4}
\usepackage{graphicx}% Include figure files
\usepackage{bm,color}% bold math
\usepackage{mathrsfs}
\usepackage{verbatim}
\usepackage{braket}

\definecolor{labelkey}{cmyk}{.4,.2,0,0}

\def\dip{U}
\def\omw{\omega_{\mbox{\tiny MW}}}
\def\pnstat{p_n^{\mbox{\tiny stat}}}
\def\pnans{p_n^{\mbox{\tiny ans}}}
\newcommand{\Tr}{\operatorname{Tr}}

\newcommand*{\GtrSim}{\smallrel\gtrsim}

\makeatletter
\newcommand*{\smallrel}[2][.8]{%
  \mathrel{\mathpalette{\smallrel@{#1}}{#2}}%
}
\newcommand*{\smallrel@}[3]{%
  \sbox0{$#2\vcenter{}$}%
  \dimen@=\ht0 %
  \raise\dimen@\hbox{%
    \scalebox{#1}{%
      \raise-\dimen@\hbox{$#2#3\m@th$}%
    }%
  }%
}
\makeatother

\newcommand{\ii}{\imath}

\graphicspath{{figures/}}

\begin{document}

\title{Evidences of spin-temperature in Dynamic Nuclear Polarization: 
an exact computation of the EPR spectrum}

\author{Filippo Caracciolo}
\author{Marta Filibian}
\author{Pietro Carretta}
\affiliation{University of Pavia, Department of Physics, Via Bassi 6, 27100-Pavia, Italy}
\author{Alberto Rosso}
\author{Andrea De Luca}
\affiliation{LPTMS, CNRS, Univ. Paris-Sud, Universit\'e Paris-Saclay, 91405 Orsay, France}

\begin{abstract}
In dynamic nuclear polarization (DNP) experiments, the compound is driven out-of-equilibrium
by microwave (MW) irradiation of the radical electron spins. 
Their stationary state has been recently probed via electron double resonance (ELDOR) techniques showing, at low temperature, 
a broad depolarization of the electron paramagnetic resonance (EPR)
spectrum under microwave irradiation. In this theoretical
manuscript, we develop a  numerical method  to compute exactly the EPR spectrum in presence of dipolar interactions.  
Our results reproduce the observed broad depolarisation and provide a microscopic 
justification for spectral diffusion mechanism. We show the validity of the spin-temperature approach 
for typical radical concentration used in dissolution DNP protocols. In particular once the interactions are properly taken into account, the
spin-temperature  is consistent with the
non-monotonic behavior  of  the EPR spectrum with a wide minimum around the irradiated frequency.
\end{abstract}

\maketitle

\section{Introduction}
Nuclear Magnetic Resonance (NMR) allows to investigate the time evolution of the nuclear magnetization in the presence of a static magnetic field.
The net magnetization per unit volume, and thus the available NMR signal, is 
proportional 
to the population difference between adjacent nuclear Zeeman levels. Being the 
energy separation 
between such levels very small with respect to thermal energy, only few spins 
contribute to the signal
which is therefore usually weak. Hence NMR spectroscopy is effective 
only at sufficiently
high nuclear spin concentrations.

Many scientific efforts have been made to overcome this sensitivity limitation, 
leading to the so-called ‘hyperpolarization’ methods. The general concept 
behind these experimental approaches is to force all the nuclear spins of a 
given sample to stay on a single Zeeman level, in order to maximize the 
population difference between the levels. The most promising strategy nowadays is 
known as Dynamic Nuclear Polarization (DNP) \cite{Wenckebach1974, 
Abragam1982a}. In the DNP protocol, 
the sample is doped with free radicals, i.e. molecules with unpaired 
electrons. It is then subject to a magnetic field and microwave irradiated. 
In absence of microwaves, the electron spins are much more polarized
than the nuclear ones, as the electron gyromagnetic ratio is 
thousands times larger than the nuclear one. Instead, when the microwaves are turned on,
at a frequency $\omw$ close to the electron Larmor frequency $\omega_e$, 
the interacting system of electrons and nuclei organizes itself in a new out-of-equilibrium steady state characterized by a strong hyperpolarization of the nuclear spins.

In the last decade, DNP has allowed the achievement of impressive results in many 
areas of science, ranging from analytical applications~\cite{Gerfen1995} to 
the development of novel diagnostic methods~\cite{Golman2006}. In 2003, 
Ardenkjaer-Larsen and co-workers developed a method to rapidly dissolve a sample, hyperpolarized at about $1$ Kelvin, 
in a solvent at room temperature~\cite{ardenkjaer2003increase}. Many research groups worldwide are currently 
exploring the novel diagnostic 
scenarios emerging from the use of hyperpolarized agents as metabolic markers.

The enhancement of the nuclear polarization emerges in the framework of a 
correlated quantum system far from equilibrium and in weak thermal contact with 
the lattice. Different DNP mechanisms can be specified according 
to the typical parameters of the system. 
In particular, the chemical shift and the $g$--factor anisotropies are responsible for the 
presence of local random magnetic fields which introduce a spread on the 
resonance frequency of nuclei and electrons, respectively. While the effect on the nuclei is 
small and, in practice one can reasonably assume that all resonate at the same frequency $\omega_n$, 
the effect on the electrons can be more significant. Accordingly the electron spectrum, 
measured in electron paramagnetic resonance (EPR) experiments, shows as characteristic features a central frequency $\omega_e$ 
and a non-negligible width $\Delta\omega_e$, depending on the local anisotropy. 
When the nuclear frequency is much larger than the electron linewidth, i.e. $\omega_n \gg 
\Delta\omega_e$, the main mechanism for nuclear polarization is a two-particles 
process, known as Solid Effect (SE). It proceeds via microwave-assisted 
forbidden transitions involving simultaneous flip-flops of one electron and one 
nucleus. In this case, the polarization transfer from electrons to nuclei occurs 
when the system is irradiated outside the EPR spectrum at a frequency $\omega_e 
\pm \omega_n$. 
When the nuclear frequency is much smaller than the electron
linewidth $\omega_n \ll \Delta\omega_e$, the polarization transfer 
from the electrons to nuclei occurs when the the EPR 
spectrum is effectively irradiated. In this case,
the simplest process inducing 
hyperpolarization involves two electrons and 
is called cross effect~\cite{hwang1967phenomenological}.
When, instead, these multi-spin resonances starts to involve many electrons, 
one expects the {\em thermal mixing} regime, typically achieved in biomedical applications. 
Its main experimental signature is that the different nuclear species in the 
compound ($^{13}$C, $^{15}$N, $^{89}$Y, $\ldots$) cool down at the same spin-temperature, 
which is
much lower than the lattice one~\cite{lumata2011dnp, kurdzesau2008dynamic}.

The hypothesis of a common spin-temperature relies on experimental observations and 
allows for explicit calculations~\cite{C3CP44667K, Serra2012, SerraColombo2013, ColomboSerra2014}, 
but is phenomenological in nature and its microscopic origin 
is still rather controversial~\cite{Hovav2013}.
With a renewed interest for DNP applications an important theoretical effort has 
been devoted to the understanding of DNP mechanisms at the microscopic level. 
An important information, 
potentially accessible through experiments, is provided 
by the EPR spectrum under irradiation, which reflects multiple properties 
of the stationary state of the spin system as a whole. 

In absence of interactions, as discussed in Sec.~\ref{sec:calc}, the EPR signal $f(\omega \simeq \omega_i)$ 
is proportional to the polarization $P_z^i$ of the $i$-th electron with a Zeeman gap $\hbar \omega_i$. 
The effect of microwave irradiation 
is encoded in the time-dependent Hamiltonian
\begin{equation}
 \label{hammw}
\hat H_{\mbox{\tiny MW}}(t) = 2 \hbar \omega_1 \sum_i \hat S_x^i \cos(\omw t)
\end{equation}
where $\omega_1$ is the intensity of the microwave field and $\omw$ its 
frequency. Then, in the rotating-wave approximation,
the electron polarization $P_z^i$ is obtained 
in terms of the solution $P_z^{\mbox{\tiny Bloch}}(\omega_i)$ of the celebrated Bloch equations, which leads to
 \begin{equation}
\label{blocheq}
P_z^{\mbox{\tiny Bloch}}(\omega) = \frac{(1+T_2^2 (\omega-\omw)^2) P_0}
{1+ T_2^2 (\omega-\omw)^2+2 T_1
   T_2 \omega _1^2} \;.
\end{equation}
Here, $P_0 = - \tanh \beta \hbar \omega/2$ is the equilibrium polarization at the lattice temperature $\beta^{-1}$
while $T_1$ and $T_2$ are respectively the spin-lattice and the spin-spin relaxation times.
In practice, the polarization of the irradiated electrons is saturated 
(i.e. $P_z^{\mbox{\tiny Bloch}}(\omw) \approx 0$), while the non-irradiated ones ($|\omega - \omw|\gg \omega_1 \sqrt{T_1/T_2}$) remain highly polarized $P_z^{\mbox{\tiny Bloch}}(\omega) \approx P_0$. 
This is the so-called \textit{hole burning} of the EPR spectrum.

On the other hand, the description based on the spin-temperature approach 
assumes that dipolar interactions induces a quasi-equilibrium behavior in the driven electron-spin system.
The traditional approach due to Provotorov~\cite{Provotorov1962}
and Borghini~\cite{Borghini1968} retains the quasi-equilibrium behavior but neglects any role
of dipolar interaction in the EPR spectrum. In this non-interacting limit,
the EPR spectrum $f(\omega)$ is then proportional to the electron polarization $P_z^i = P_z^{\mbox{\tiny Borg}}(\omega_i)$, 
with 
\begin{equation}
\label{PzST}
P_z^{\mbox{\tiny Borg}}(\omega) = - \tanh\Bigl(\frac{\beta_s \hbar (\omega - \omega_0)}{2} \Bigr). 
\end{equation}
This expression depends on two intensive parameters\footnote{This approach is often presented in the literature employing, rather than the frequency 
$\omega_0$, the parameter $\alpha$ with the dimension of an inverse temperature conjugated to the 
Zeeman component of the energy. It is related to our intensive parameter $\omega_0$ by
$$\alpha=  \frac{\beta_s (\omega_e - \omega_0)}{\omega_e}$$
}: the inverse spin-temperature 
$\beta_s$, which can be very different from the inverse lattice temperature $\beta$ 
and the effective magnetic field $\omega_0$, which is usually close to $\omw$.

The most counterintuitive feature of $P_z^{\mbox{\tiny Borg}}(\omega)$ 
is the monotonic behavior with 
a change of sign for $\omega = \omega_0$ so that
for a wide range of frequencies, the electron spins are aligned parallel 
to the external magnetic field. 
Such an electron-polarization inversion was indeed observed long-time ago in the irradiated EPR spectrum of the
$\mbox{Ce}^{3+}$ ions 
in a $\mbox{CaWO}_4$ crystal~\cite{Atsarkin1970}. Recently, 
a set of experiments have been performed
at several temperatures and microwave 
intensities~\cite{schosseler1994pulsed, granwehr2008multidimensional, hovav2015electron}, but
none of them observed this characteristic inversion.
This fact has been used as an evidence invalidating the spin-temperature description, even 
at low-temperature, where dissolution DNP is efficiently employed.
However, in this regime, an anomalously large hole burning is observed: the EPR spectrum 
displays an important depolarization throughout its full width, which is inconsistent
with the behavior of $P_z^{\mbox{\tiny Bloch}}(\omega)$.
To account for these experimental results, the authors of Ref.~\cite{hovav2015electron} 
introduced a system of rate-equations for the electron polarization.
This model contains a phenomenological term describing the flip-flop transition between pair of electron spins 
resonating at different frequencies. Such a term is supposed to describe the
electron spectral diffusion and the broadening of the hole burning, but does not have a
clear microscopic origin.

In this manuscript, we present an exact microscopic calculation of the EPR spectrum $f(\omega)$
which goes beyond the non-interacting approximation $f(\omega_i) \propto P_z^i$. 
For low concentrations, the microwaves dig a narrow hole in the EPR spectrum which is consistent
with the simple behavior of $P_z^{\mbox{\tiny Bloch}}(\omega)$.
Instead, for higher concentration, a large reorganization of the spectrum is observed, which 
we will show to perfectly agree with the spin-temperature description of the interacting model. Nonetheless, 
the EPR spectrum can be different from the non-interacting Borghini limit in Eq.~\eqref{PzST}. 
In particular, for moderate microwave intensity,
the fingerprint of spin-temperature is a broad depolarization, without any polarization inversion, 
similar to the one observed in 
the recent low-temperature experiments~\cite{hovav2015electron}.
Only for strong microwave irradiation, this reorganization
displays the polarization inversion predicted by Eq.~\eqref{PzST}. 
Note that our results are obtained for the EPR spectrum 
without assuming any macroscopic electron spectral diffusion.

The paper is organized as follows. In Sec.~\ref{sec:model} 
we review the model introduced in~\cite{DeLuca2015,de2016thermalization} 
for the electron spin system and provide the details for the numerical implementation. In 
Sec.~\ref{sec:calc}, we derive an explicit formula given in Eq.~\eqref{eprformula}, 
for the EPR spectrum. Note that this formula is exact, and does not reduce $f(\omega)$ to the individual electron
polarizations $P_z^i$. The procedure to test the validity of the spin-temperature 
concept and the comparison with numerical data are given in Sec.~\ref{sec:results}. 
In the conclusion, we summarize our main result: 
the electron spectral diffusion of the EPR spectrum is induced by the dipolar interactions
and can be described by our microscopic model; moreover
the presence of a strong electron spectral diffusion is a signal
of a quasi-equilibrium behavior in the driven stationary state. 
We also comment on how the polarization performance 
is influenced by the spatial arrangement of the radicals in samples used for \textit{in vivo} metabolic imaging.

\begin{figure}[ht]
\centering
  \includegraphics[height=5cm]{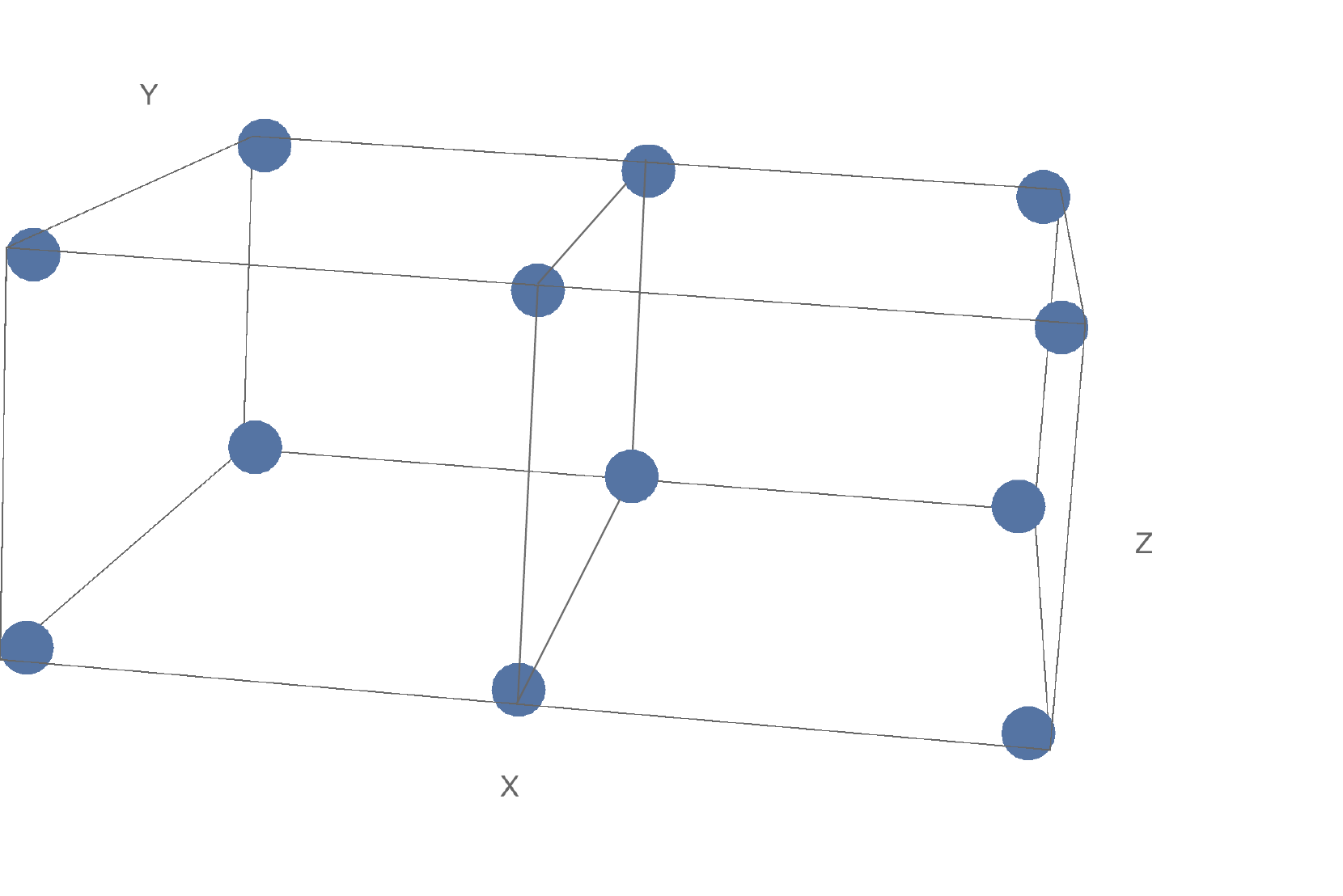}
  \caption{Cubic lattice representing the spatial arrangement of the $N=12$ 
electron spins. The  lattice spacing is $a=0.1 
(C \cdot N_a)^{-1/3}$ (for $ C = 1.5 \times 10^{-3}$ M, $a = 114 \;  \AA$, for $C = 
15\times 10^{-3}$ M, $a= 54 \; \AA$), with $C$ the molar radical concentration and $N_a$ the Avogadro number. Here periodic boundary condition are 
implemented 
  and the distance between two spins is defined as the length of the shortest 
path
  so that each spin has $4$ spins at distance $a$, $5$ spins at 
distance $\sqrt{2} a$, 
  $2$ spins at distance $\sqrt{3}a$.}
  \label{fgr:cube}
\end{figure}

\begin{table}[h]
\small
  \caption{Parameteres of the simulation: $T_{1}$ and  $T_{2}$ are respectively the 
  longitudinal and transverse electron relaxation times, $\omega_{e} $ is the electron Larmor frequency, 
  $\Delta \omega_{e}$ the electron linewidth in absence of dipolar 
interaction and $\beta^{-1}$ is the lattice temperature.}
  \label{tbl:filippo}
  \begin{tabular*}{\columnwidth}{@{\extracolsep{\fill}}llllll}
    \hline
   $T_{1}$ (s)& $T_{2}$ (s) & $\omega_{e}$ (GHz)& $\Delta \omega_{e}$ (GHz) & 
$\omw$  (GHz)& $\beta$ (K$^{-1}$)  \\
    \hline
    $1$ & $10^{-6}$ &$93.9$&0.108 & $93.8685$ &   $0.04$ \\
    \hline
  \end{tabular*}
\end{table}

\section{Review of the model and numerical procedure \label{sec:model}}
We review the model introduced in \cite{DeLuca2015}, employed in the description of the electron spins,
under microwave irradiation, 
We consider a collection of $N$ electron spins described by the Hamiltonian
\begin{equation}
\label{zeeman}
 \hat H_{S} = \sum_{i =1}^N \hbar \left( \omega_e +\Delta_i \right) \hat S_z^i 
 %- \omega_n \sum_{j=1}^{N_n} \hat I_z^j 
 + \hat H_{\mbox{\tiny dip}}.
\end{equation} 
The $\Delta_i$'s label the inhomogeneous field due to the $g$-factor anisotropy.
Here, we denote the spin $1/2$ operator on $i$-th electron with $\hat 
S_\alpha^i$ for $\alpha = x,y,z$. 
For large magnetic fields, the dipolar interactions can be treated as a 
perturbation of Zeeman energy in \eqref{zeeman}. 
This leads to the secular approximation~\cite{Abragam1982a}
\begin{equation}
  \label{secdipolar}
  \hat H_{\mbox{\tiny dip}} = \sum_{i<j} \dip_{ij} \left[4 \hat{S}^i_z 
\hat{S}^j_z -(\hat{S}^i_+ \hat{S}^j_- + 
  \hat{S}^i_- \hat{S}^j_+)\right]
 \end{equation}
 where $\dip_{ij} = \mu_{0} \hbar^2 \gamma_{e}^2(1 - 3 \cos^2\theta_{ij}) 
/(16\pi|\mathbf{r}_{ij}|^{3})$, $\mu_{0} $ is the vacuum magnetic permeability, 
$\gamma_{e}$ is the electron gyromagnetic ratio 
 and $\hbar \mu_{0} \gamma_{e}^2/16\pi$ is $ 81.7 \times 2\pi$ GHz $\AA^{3}$. 
 Here, $\theta_{ij}$ is the angle between the field (taken along $z$) and 
$\mathbf{r}_{ij}$, the vector connecting the $i$-th and the $j$-th spin. 
$[\hat S_{+}^i, \hat S_{-}^i] = 2 \hat S_z^i$, with $\hat S^i_\pm = \hat 
S^i_x \pm i \hat S^i_y$.

Concerning the microwaves in Eq.~\eqref{hammw},
we can assume that $\omega_1$ is few tens of KHz, remaining therefore much 
weaker than the other terms in the Hamiltonian.
Thus at frequency $\omw$ in the rotating frame 
\begin{equation}
\rho^{\mbox{\tiny rot}} \equiv e^{\ii \hat S_z \omw t} \rho e^{-\ii \hat S_z 
\omw t}
\end{equation}
where $\rho$ is the density matrix of the electron spin system in the laboratory frame and 
$\rho^{\mbox{\tiny rot}}$ is the one in the rotating frame.
Then, in the evolution equation for $\rho^{\mbox{\tiny rot}}$, one can neglect 
the fast oscillating terms induced by microwaves and restrict to the 
dominant
one which is time-independent. We arrive at the Liouville equation
\begin{equation}
\label{liouvillerot}
 \frac{d}{dt}\rho^{\mbox{\tiny rot}} = - \frac{\ii}{\hbar} [ \hat 
H^{\mbox{\tiny rot}}, \rho^{\mbox{\tiny rot}}]
\end{equation}
where the Hamiltonian, including the microwave irradiation,
in the rotating frame takes then the form
\begin{equation}
 H^{\mbox{\tiny rot}} = \hat  H_{S} - \hbar \omw \sum_i \hat S_z^i  + \hbar 
\omega_1 \sum_i \hat S_x^i
\end{equation}

\subsection{The master equation in the Hilbert approximation}
The Liouville equation introduced in Eq.~\eqref{liouvillerot} has to be 
modified in order to take into account
the spin-lattice relaxation mechanisms. The resulting dynamics describe the 
evolution of $N$ electron spins and involves a linear system with $4^N$ 
components. This strongly limits the accessible system 
sizes. An important simplification occurs
in our case, as the spin-spin relaxation times ($T_2$) are much faster 
than the spin-lattice ones ($T_1$). 
Using the approach derived in \cite{hovav2010theoretical, Hovav2012, Hovav2013, 
DeLuca2015}, 
the time-evolution of the spins, in the so-called Hilbert approximation, 
reduces 
to the evolution of diagonal elements of the $\rho^{\mbox{\tiny rot}}$ in the 
basis of eigenstates $\ket{n}$ of $\hat H_S$. 
We obtain then a classical master equation for the probability $p_n \equiv 
\rho^{\mbox{\tiny rot}}_{nn}$ 
of occupying the eigenstate $\ket{n}$ 
(i.e. $\hat H_S \ket{n} = \epsilon_n \ket{n}$):
\begin{equation}
 \label{mastereq1}
\frac{dp_n}{dt} = \sum_{n' \neq n}  W_{n' \to n} p_{n'} - W_{n\to n'} p_n \;.
\end{equation}
The transition rate between the pair of eigenstates $\ket{n}, \ket{n'}$ 
has the form $W_{n,n'} = W^{\text{latt}}_{n, n'} + W^{\text{MW}}_{n, n'}$, with
\begin{align}
\label{matrixtransitionBATH}
W^{\text{latt}}_{n, n'} &=  \frac{2h_\beta(\Delta \epsilon_{n,n'}) 
}{T_1}\sum_{j=1}^N \sum_{\alpha=x,y,z} |\bra{n}\hat 
S_{\alpha}^{j}\ket{n'}|^2\;,\\
\label{matrixtransitionMW}
W^{\text{MW}}_{n, n'} &= \frac{4 \omega_1^2 T_2 |\bra{n}  \sum_{j=1}^{N} \hat 
S_{x}^{j}  \ket{n'}|^2}{1+ T_2^2 (|\epsilon_n - \epsilon_{n'}|/\hbar - \omw 
)^2}\;.
\end{align}
Eq.~\eqref{matrixtransitionBATH} contains spin-flips induced by spin-lattice 
relaxation mechanism 
on a time scale $T_{1}$ and
the function $h_\beta(x)=e^{\beta x}/(1+ e^{\beta x})$ 
assures the detailed balance and convergence to Gibbs equilibrium at the lattice
temperature $\beta^{-1}$.
Eq.~\eqref{matrixtransitionMW} encodes the effect of microwaves, which, as expected, are 
particularly effective for transitions
under the resonance
condition: $\epsilon_n - \epsilon_{n'} \simeq \hbar \omw$. The time-scale $T_2$ 
is identified with the electron transverse relaxation time, while the 
time-scale $T_1$
with the electron spin-lattice relaxation time.

\subsection{Numerical implementation \label{sec:num}}
The stationary state  of Eq.~\eqref{mastereq1} is obtained numerically for a 
system of $N=12$ electron spins. In the experimental practice, the radical molecules
are dissolved in a frozen amorphous mixture~\cite{amorphous2013} as 
a substantial decrease of the polarization is observed 
in samples prepared in a crystalline phase. Here, for simplicity, we
disposed, in our simulations, the electron spins on the cubic lattice, as shown in Fig.\ref{fgr:cube}. 
We postpone the discussion about 
the importance of the spatial arrangement of the radicals.

The inhomogeneous magnetic fields, $\Delta_i$, are derived from the gaussian 
distribution centered at $\omega_{e}$ and with 
standard deviation $\Delta \omega_e$. We  diagonalize the hamiltonian $ \hat H_{S}$ 
and compute the $2^{N}$ eigenstates $\ket{n}$ of energy $\epsilon_n$ and total 
electron magnetization 
$s_{z,n}= \bra{n} \hat S_z \ket{n}$ with $\hat S_z = \sum_i \hat S_z^i$. 
Then, the rates in (\ref{matrixtransitionBATH},~\ref{matrixtransitionMW}) can 
be computed using the matrix 
elements between each pair of eigenstates and the parameters in Table \ref{tbl:filippo}. 
This set of parameters is chosen to represent a pyruvic acid sample doped with trityl radical 
at the temperature $\beta^{-1} = 1.2$ K and an external magnetic field of $3.35$ T. These conditions
have been studied experimentally in great detail~\cite{johannesson2009dynamic, macholl2010trityl, filibian2014role} 
and represent a good test for our theoretical model because 
of clear evidences of thermal mixing in nuclear polarizations. 

The occupation probabilities in the stationary state $\pnstat$ are finally 
obtained 
setting $dp_n/dt = 0$ Eq.~(\ref{mastereq1}) and solving the resulting 
linear system. This procedure is repeated over many realizations of the 
inhomogeneous fields. The EPR spectrum presented in this work are averaged over 
$\sim 600$ realizations (see below for the details of the averaging).

\begin{figure*}[ht]
\begin{center}
\includegraphics[width=0.97\columnwidth]{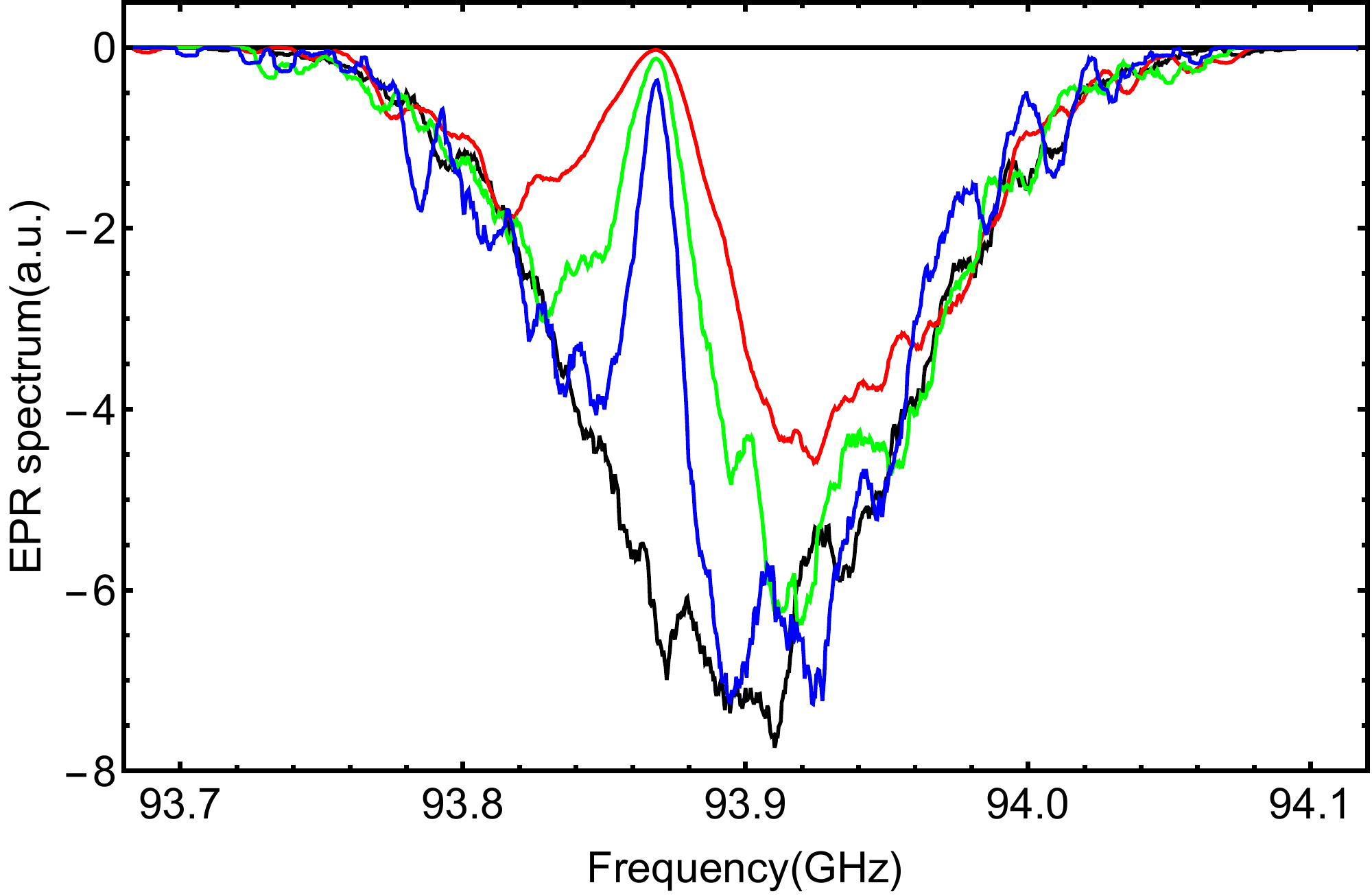} 
\includegraphics[width=0.97\columnwidth]{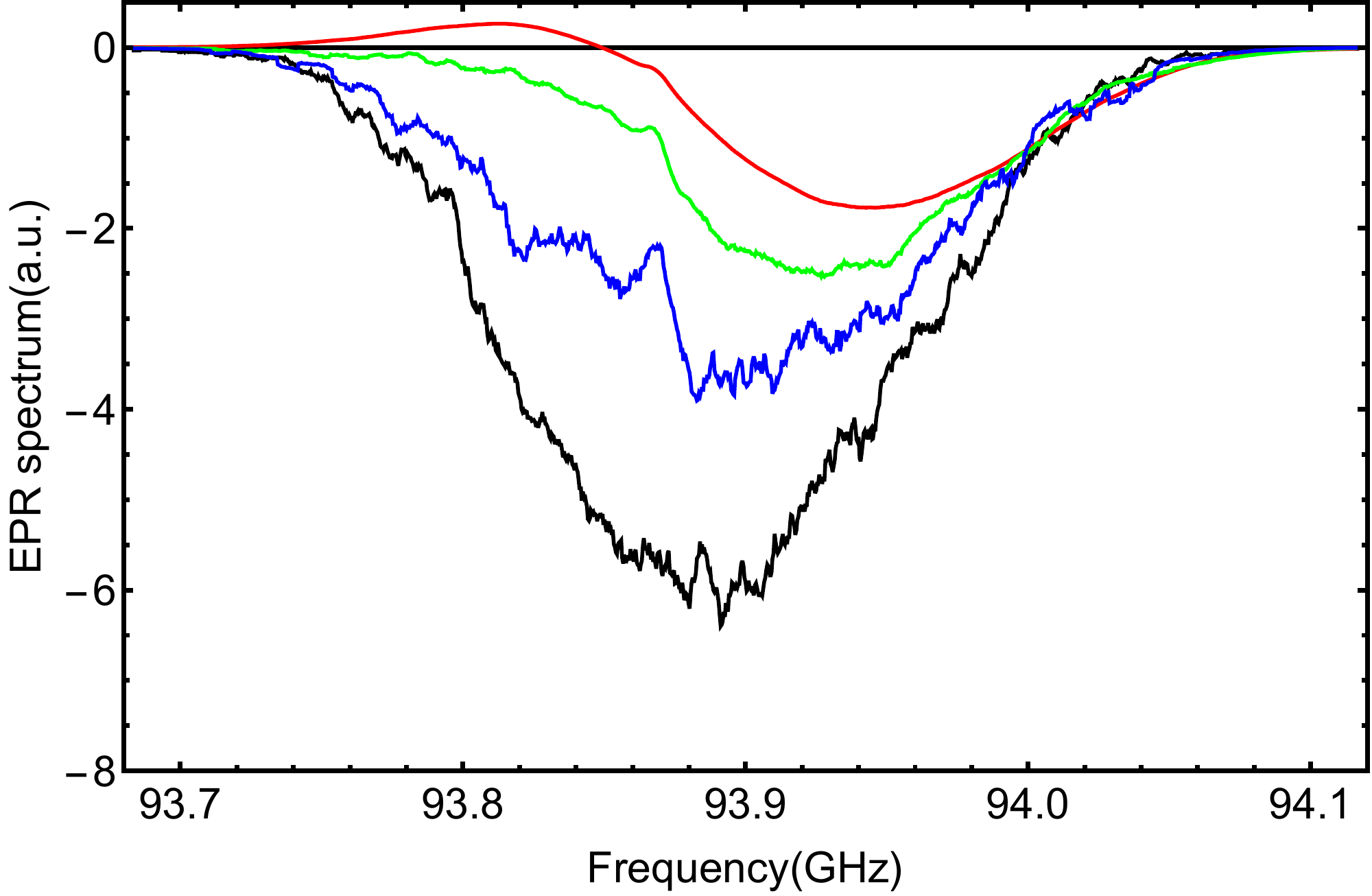}
\caption{Numerical EPR spectrum (from exact diagonalization of the density matrix) $f(\omega)$ under MW irradiation for the two radical concentrations of $1.5$ mM (left) 
and $15$ mM (right) as
obtained from Eq.~\eqref{eprformula} with $p_n = \pnstat$. 
Several MW intensities are considered, from the bottom to the top: $\omega_{1}=0 
$ GHz,$\omega_{1}=0.625 \times 10^{-5} $ GHz,
$\omega_{1}=0.125 \times 10^{-4}$ GHz, $\omega_{1}= 0.25 \times 10^{-4} 
$ GHz. 
Low concentration: the irradiated profile displays a hole burning shape; no 
inversion of the polarization is observed.
Large concentration:even in presence of weak irradiation, the EPR profile 
entirely reorganizes 
but it shows the striking inversion of the 
polarization only at high MW intensity. 
\label{Fig:comparisonMW} 
}
\end{center}
\end{figure*}

\begin{figure*}[!]
\centering
\includegraphics[width=\columnwidth]{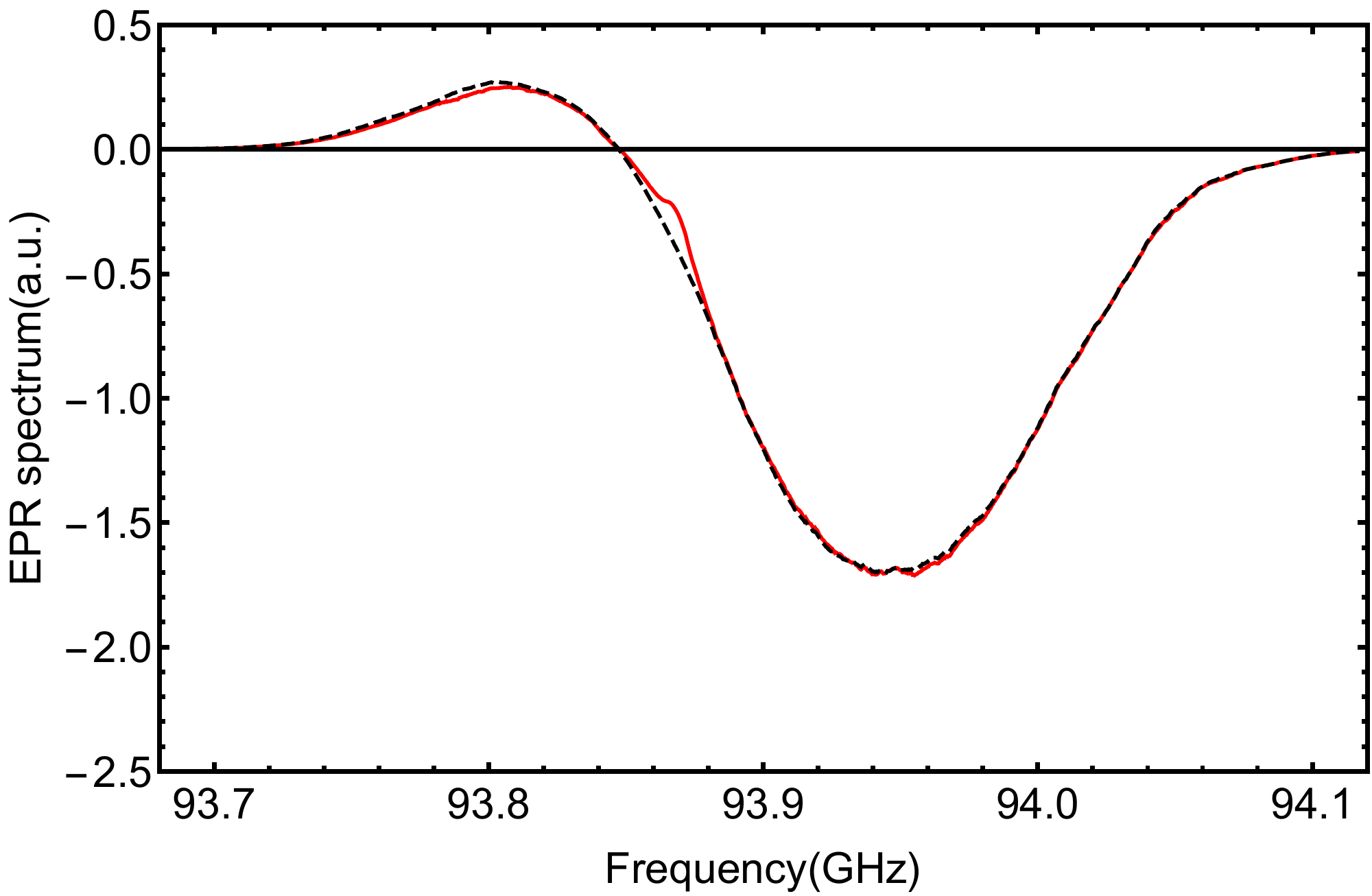}
\includegraphics[width=\columnwidth]{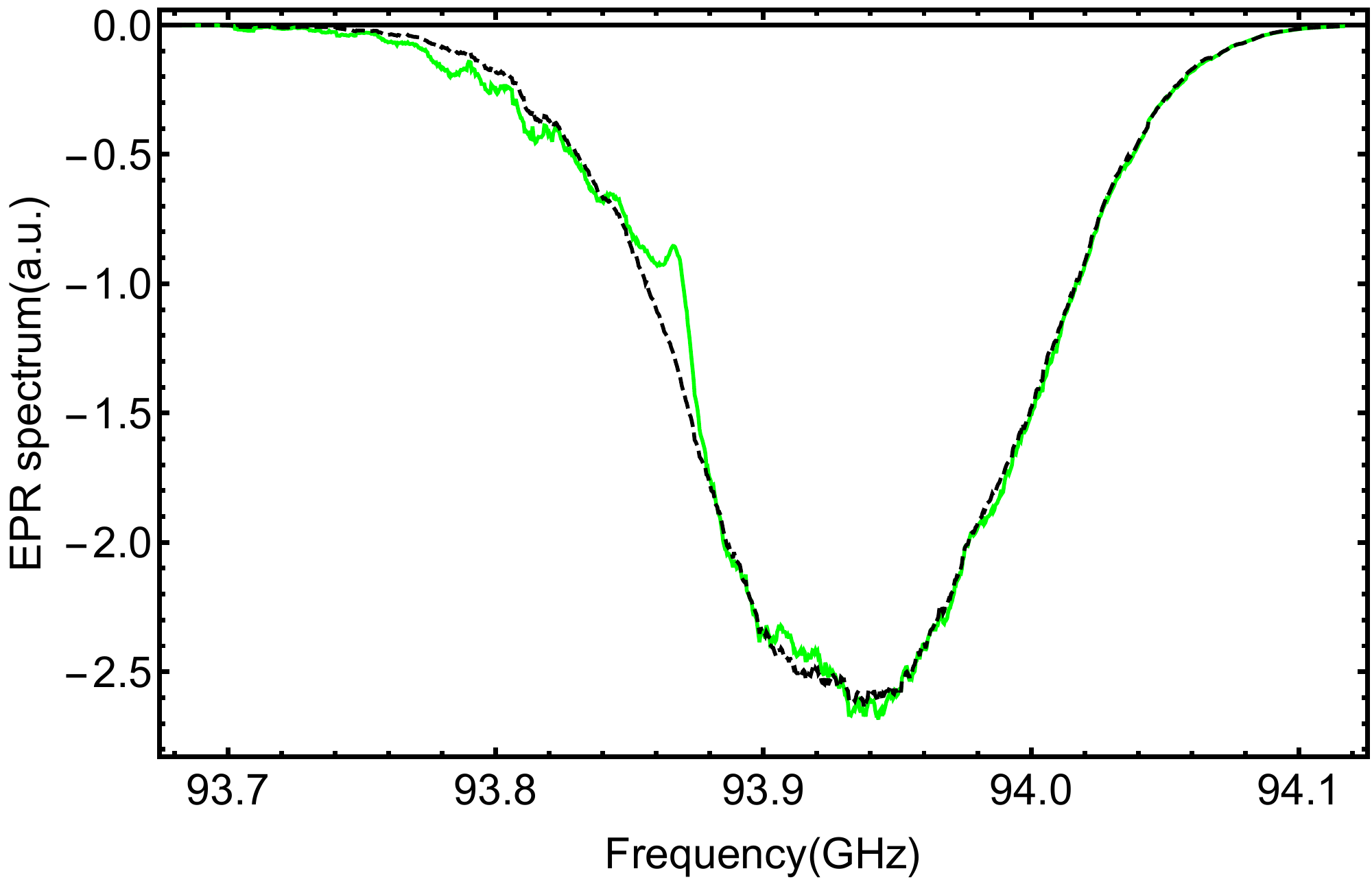}
\caption{Comparison between the numerical EPR spectrum with $p_n = \pnstat$ (continous line)
with the one obtained through the spin-temperature approach $p_n = \pnans$ in 
Eq.~\eqref{stans} (dashed line) at high concentration C=
$15$ mM. Left: MW intensity $\omega_{1}=0.25 \times 10^{-4}$ GHz. Right: 
$\omega_{1}=0.125 \times 10^{-4}$ GHz.
We observe a perfect agreement between the two curves in both cases, even in 
absence of polarization inversion (right).
\label{Fig:15mM}}
\end{figure*}

\section{Exact EPR spectrum formula in presence of interaction\label{sec:calc}}
In this section, we derive an exact expression for the EPR spectrum 
in presence of dipolar interactions between the electron spins. 
In a pulsed EPR experiment, one applies a $\pi/2$ pulse, 
which flips the longitudinal magnetization along $z$ in the $xy$ plane. For 
simplicity,
we assume that the final magnetization is along the $y$-axis:
$\hat U_{\pi/2} \hat S_z^i \hat U_{-\pi/2} = \hat S_y^i$, 
where $\hat U_{\theta} = e^{i \theta \hat S_x}$ is a rotation of angle $\theta$ 
around the $x$-axis and $\hat S_x = \sum_i \hat S_x^i$.
Thus, such a pulse induces an abrupt change for the density matrix $\rho$
\begin{equation}
 \rho \longrightarrow \rho_{\pi/2} \equiv \hat U_{\pi /2 } \rho \hat U_{-\pi/2},
\end{equation}
where the subscript $\pi/2$ indicates quantities computed after the pulse.
One can easily check that the original longitudinal magnetization of the spin 
$i$
is flipped along the $y$-axis: 
\begin{equation}
P_z^i = 2 \Tr[\hat S_z^i \rho] \longrightarrow P_{y,\pi/2}^i , 
\end{equation}
where $P_{\alpha, \pi/2}^i = 2\Tr[\hat S_\alpha^i \rho_{\pi/2}]$ is the polarization in the $\alpha$-axis,
defined as the magnetization along $\alpha$ and normalized between $-1$ and $1$. 
The polarization in the $xy$-plane is usually dubbed ``coherence'' and 
is a property of the system detected in a magnetic resonance experiment.
After the pulse, the polarization of the spin $i$ in the $xy$-plane rotates at 
the Zeeman frequency 
$\omega_i = \omega_e + \Delta_i$.
Moreover, a spin-spin dephasing is 
induced by dipolar interactions 
with the other electron spins.
This can be modeled as an effective exponential decay of the longitudinal 
polarization with a characteristic time $T_2$:
\begin{equation}
\begin{bmatrix}
P_{x,\pi/2}^i(\tau) \\ P_{y,\pi/2}^i(\tau)
\end{bmatrix}
= \begin{bmatrix}
- \sin(\omega_i \tau) \\ \cos(\omega_i \tau)
\end{bmatrix}
P_z^i(\tau=0) e^{-\tau/T_2} \;.
\end{equation}
Then, in this effective non-interacting picture, one can obtain the contribution of the $i$-th spin to the 
EPR spectrum as
\begin{equation}
\label{eprdef}
 f_i (\omega) = \operatorname{Re}\left[\int_{0}^\infty \frac{dt}{\pi} g_i(\tau) 
e^{- i \omega \tau}\right] = 
\frac{T_2 P_z^i/\pi}{T_2^2 (\omega - \omega_i)^2 + 1},
\end{equation}
where we introduce the function
\begin{equation}
\label{gdef}
 g_i(\tau) \equiv P_{y,\pi/2}^i(\tau) - \ii P_{x,\pi/2}^i(\tau)
% 
% - 2 \ii \Tr[ \hat S_+^i(\tau) \rho_{\pi/2}] \;.
\end{equation}
and used that, before the pulse $(\tau < 0)$, the magnetization in the 
$xy$-plane vanishes.

In this manuscript, we treat explicitly the effect of dipolar interactions
between spins and the parameter $T_2$ only accounts for the microwave strength, as given in Eq.~\eqref{matrixtransitionMW}. In particular, at time much shorter than $T_1$, 
the time evolution is simply governed by the quantum Hamiltonian $\hat H_S$ in 
\eqref{zeeman}.
Then, the polarization of the spin $i$ takes the form
\begin{equation}
 P_{\alpha,\pi/2}^i(\tau) = 2\Tr[\hat S_\alpha^i(\tau) \rho_{\pi/2}]
\end{equation}
where we chose the Heisenberg picture for the time-evolution, i.e. 
$\hat S_\alpha^i(\tau) = e^{\ii \hat H_S \tau} \hat S_\alpha^i e^{-\ii \hat H_S 
\tau} $.
Therefore, the function $g_i(\tau)$ introduced in Eq.~\eqref{gdef} 
can be rewritten as
\begin{equation}
 \label{gdef1}
g_i(\tau) = - 2 \ii \Tr[ \hat S_+^i(\tau) \rho_{\pi/2}] \;.
\end{equation}
We now note that $e^{\ii \pi \hat S_x^i/2} =  2^{-1/2} (1 + 2 i \hat S_x^i)$ 
and arrive at
\begin{align}
\label{gdefsimple}
 g_i(\tau) 
 =\Tr[\hat S_+^i(\tau)\hat S_-^i (0)\rho]- \Tr[ \hat S_-^i(0) \hat S_+^i(\tau) 
\rho]  \;.
\end{align}
which shows the function $g_i(\tau)$ is nothing else but the spin-spin 
time-correlation function.
To derive this last equation, we used $[ \hat S_z, \hat H_S ] = 0$, since
only the terms conserving the total magnetization should be retained. 

Since dephasing is fast, we can safely assume 
that the stationary density matrix before the pulse $\pi/2$ was diagonal in the 
basis of eigenstates
of $\hat H_S$ \cite{DeLuca2015,hovav2010theoretical}: 
\begin{equation}
\rho = \sum_n p_n \ket{n} \bra{n} \;.
\end{equation}
Using this fact, we rewrite Eq.~\eqref{gdefsimple} as
\begin{equation}
\label{gexpl}
g_i(\tau) = \sum_{n,m} (p_n - p_m) e^{\ii \tau (E_n - E_m)} |\bra{n} \hat S_+^i 
\ket{m}|^2 \;. 
\end{equation}
The EPR spectrum in Eq.~\eqref{eprdef} generalizes to
\begin{equation}
 \label{eprdefgen}
 f(\omega) = \frac1N\sum_i\operatorname{Re} 
 \left[\int_{0}^\infty \frac{dt}{\pi} g_i(\tau) e^{- i \omega \tau - \eta\tau} 
\right]\;.
\end{equation}
where we introduced a small cutoff $\eta>0$ to obtain a smooth spectrum and we 
averaged over the ensemble of spins.
Employing Eq.~\eqref{gexpl}, we arrive to the final expression
\begin{equation}
\label{eprformula}
f(\omega) = \frac{\eta}{N \pi}\sum_{n,m} \frac{(p_n-p_m)  \sum_i |\bra{n} S_+^i 
\ket{m}|^2}{(\omega - (\epsilon_n - \epsilon_m))^2 + \eta^2}
\end{equation}
The cutoff $\eta$ takes care of the finite-size effects:
if $\eta \to 0$, for a finite $N$, the function $f(\omega)$ 
is the sum of a set of discrete $\delta$-peaks 
in correspondence of the values $\omega = \epsilon_n - \epsilon_m$; 
however, the number of these peaks grows exponentially in $N$ and
leads to a smooth distribution $f(\omega)$ when $N\to \infty$. 
In practice, in our data, we took $\eta \to 0$, but we integrated $f(\omega)$ on 
a small 
interval of width $\delta \omega = \Delta\omega_e/150$: $\tilde f(\omega) = 
\frac{1}{\delta\omega}\int_{\omega}^{\omega + \delta\omega} f(\omega)$.
This final quantity $\tilde f(\omega)$ is then averaged over different 
realizations of the fields $\Delta_i$. For simplicity we will drop the tilde in the following.

In the non-interacting case, the expression in Eq.~\eqref{eprformula} 
would simplify to
$f(\omega) = \frac{1}{N}\sum_i P_z^i \; \delta(\omega - \omega_i)$,
which explicitly relates the electron polarizations to the EPR spectrum.
Note that, while in the interacting case, $f(\omega)$ is not directly connected to the 
electron polarizations $P_z^i$, the area below $f(\omega)$ remains equal to the total
polarization along $z$:
\begin{equation}
 \label{sumrule}
 \int_{-\infty}^\infty d\omega f(\omega) = 2 \Tr [ \hat S_z \rho] 
\end{equation}

\begin{figure}[!]
\centering
\includegraphics[width=\columnwidth]{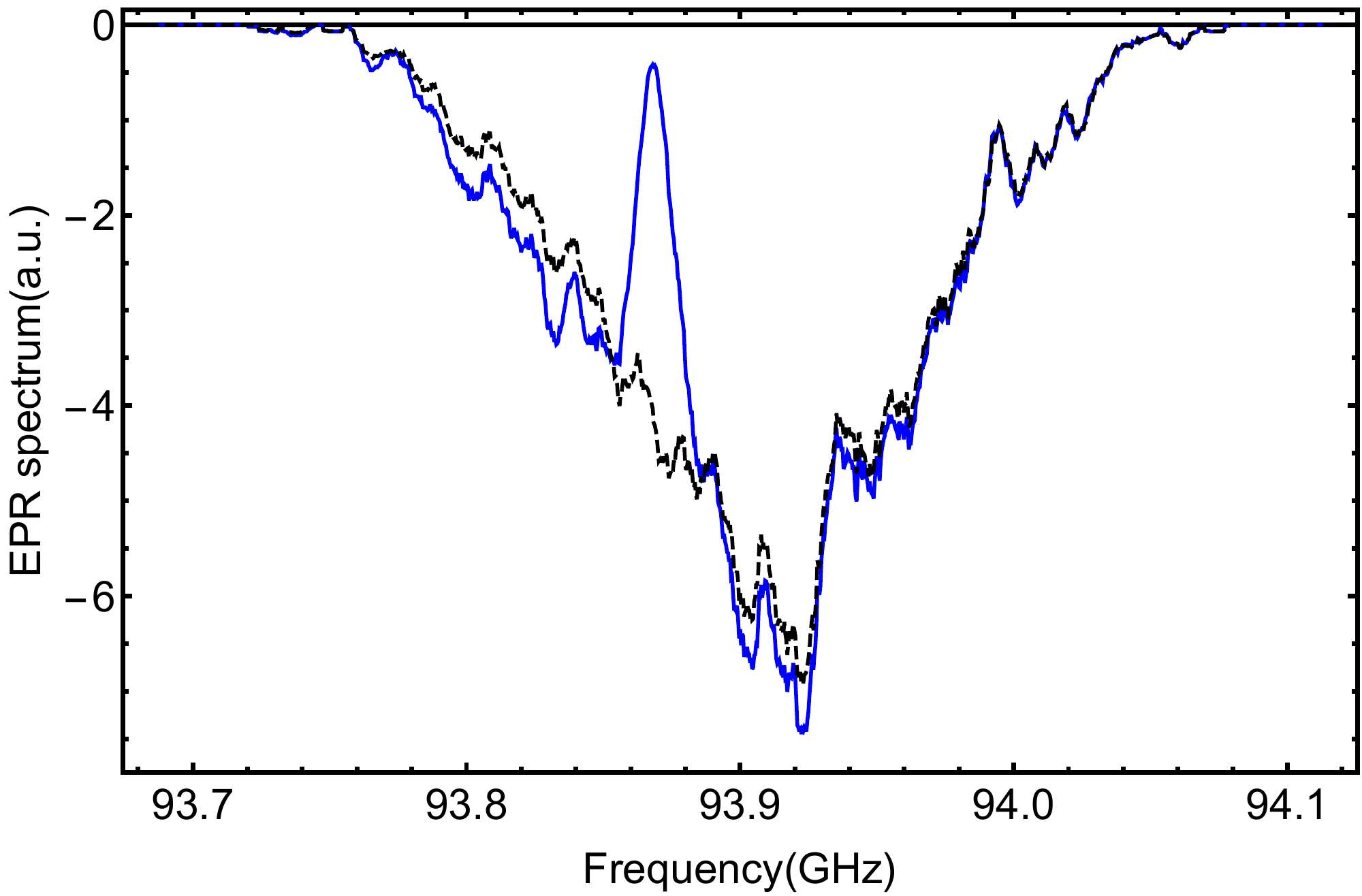}
\caption{Comparison between the numerical EPR spectrum with $p_n = \pnstat$(continous line)
with the one obtained from the spin-temperature approach $p_n = \pnans$(dashed line) in 
Eq.~\eqref{stans} for low concentration
$1.5$ mM and $\omega_{1}=0.625 \times 10^{-5}$.
We observe that the spin-temperature approach resembles the non-irradiated 
spectrum shown in Fig.~\ref{Fig:comparisonMW}
and fails to reproduce the hole-burning shape which characterizes
the stationary distribution.
}
\label{Fig1d5mM0125}
\end{figure}

\section{Results \label{sec:results}}

We considered two different radical concentrations: $C = 15 $ mM and $C = 1.5 $ mM, 
the former corresponding to a typical radical concentration used in DNP experiments, 
while the latter to a low radical concentration, which does not allow to reach sizeable nuclear polarization levels. 
In Fig.~\ref{Fig:comparisonMW},
we compare the response of the two cases for several intensities of the 
microwave irradiation. 
The EPR spectra are obtained inserting $p_n = \pnstat$, obtained as explained 
in Sec.~\ref{sec:num},
in Eq.~\eqref{eprformula}.
In absence of microwaves, the spectra, for the two radical concentrations, 
are very similar and the broadening induced by the dipolar interaction is weak.
Turning on the microwaves, the two spectra appear very different: in the 
low-concentration sample, a hole burning
modifies the spectrum around the microwave frequency $\omw$. Increasing the 
intensity of the microwave irradiation,
such an effect becomes broader and deeper. On the contrary, in the case of 
high-concentration, microwaves
affect the whole EPR spectrum even at very low intensity. In particular, for 
strong irradiation, the spectrum shows
an inversion of polarization with respect to the equilibrium signal. This 
inversion is the main manifestation of 
the spin-temperature according to the toy model of Borghini \cite{Borghini1968, 
Abragam1982a}. \textit{Nevertheless, it remains as an open question whether
in these systems, the spin-temperature could anyhow be an effective description.}
This point is addressed in the next subsection, where we discuss a general method to test
the validity of the spin-temperature approach~\cite{de2016thermalization}.

\subsection{The spin-temperature approach}
From our numerical simulation, we have the possibility to perform an explicit 
check of the validity of the spin-temperature
description. According to the spin-temperature hypothesis, one 
assumes that 
\begin{equation}
\label{stans}
 \pnstat \sim \pnans \equiv \frac{e^{-\beta_s (\epsilon_n - \omega_0 
s_{z,n})}}{\mathcal{Z}}
\end{equation}
which depends on two intensive parameters: the inverse spin-temperature 
$\beta_s$ and the
effective magnetic field $\omega_0$. The normalization $\mathcal{Z}$ is chosen 
to enforce $\sum_n \pnstat = 1$. 
These two quantities can be fixed imposing that the stationary state described
by \eqref{stans} has the same total energy and total magnetization of the true 
stationary state, leading to the two equations:
\begin{subequations}
\label{stansatz}
\begin{align}
 \langle \hat H_S \rangle = \sum_{n} \pnstat \epsilon_n &= \sum_{n} \pnans 
\epsilon_n \;,\\
 \langle \hat S_z \rangle = \sum_{n} \pnstat s_{z,n} &= \sum_{n} \pnans 
s_{z,n}\;. 
 \end{align}
 \end{subequations}
For large system sizes, we expect that the values of these two parameters do 
not fluctuate between
different realizations. However, for $N = 12$, we decided to solve these two 
equations for every single realization,
thus computing $\beta_s$ and $\omega_0$ in each case. Finally,
the EPR spectrum in the spin-temperature approach, is computed again using 
Eq.~\eqref{eprformula}
but replacing $p_n$ with $\pnans$. 
In Fig.~\ref{Fig:15mM}, we show the comparison between the real EPR spectrum 
and the one
obtained with the spin-temperature hypothesis for the sample at $15 $ mM and 
different microwave intensities. In both cases, there is a clear agreement demonstrating the validity of the spin-temperature approach for radical concentrations typically used in DNP protocols. It is important to notice that the inversion appears only for the strongest microwave irradiation. 
Finally, in Fig.~\ref{Fig1d5mM0125} we considered the spin-temperature approach 
for the low-concentration sample.
In this case, the spin-temperature approach fails to reproduce the hole burning in the spectrum and 
instead, as
the absorbed irradiation gets redistributed among the full system, 
the final result closely resemble the
non-irradiated spectrum of Fig.~\ref{Fig:comparisonMW} right.

\subsection{The non-interacting case}
For the sake of completeness, we now discuss the extreme limit of vanishing dipolar interactions.
In this case, the stationary value of the total energy and total magnetization 
can be easily computed
in terms of the electron polarizations
\begin{subequations}
\label{blochtotal}
\begin{align}
 \langle \hat H_S \rangle &= \frac{1}{2}\sum_i \omega_i P_z^i \;,\\
 \langle \hat S_z \rangle &= \frac12 \sum_i P_z^i\;.
 \end{align}
 \end{subequations}
The exact value of the electron polarizations in the stationary state is 
provided by the solution of Bloch equation introduced in Eq.~\eqref{blocheq}, with 
$P_z^i = P_z^{\mbox{\tiny Bloch}}(\omega_i)$

\begin{figure}[ht]
\centering
\includegraphics[width=\columnwidth]{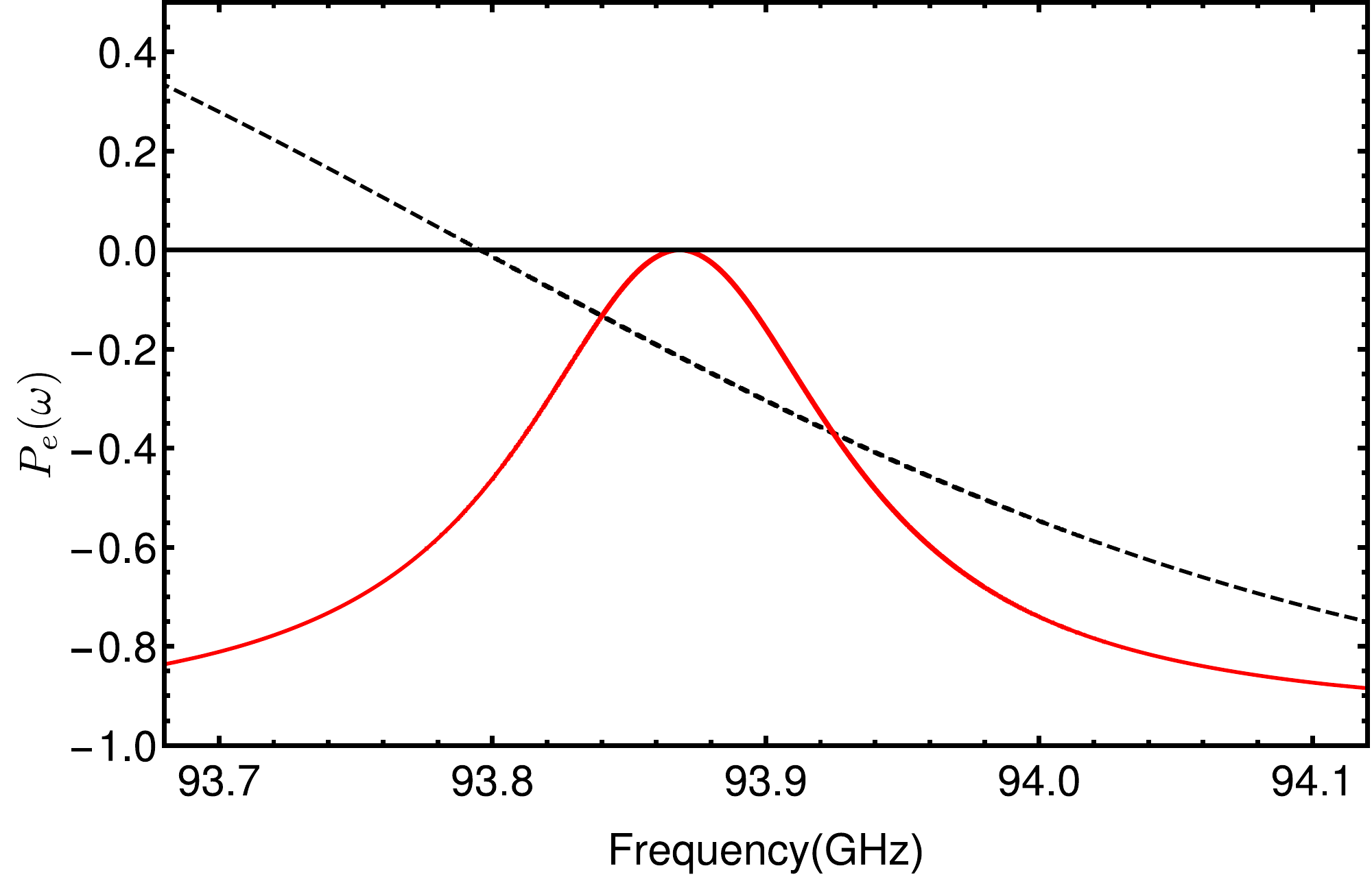}
\caption{Comparison between the electron polarizations as predicted by the 
Bloch equations
$P_e^i = P_e^{\mbox{\tiny Bloch}}$ given in Eq.~\eqref{blocheq} (which 
corresponds to the exact stationary state
in absence of dipolar interactions; solid line) with the one obtained within the 
spin-temperature apporach 
$P_e^i = \tanh(\beta_s (\omega_i - \omega_0)/2)$(dashed line). 
Both curves are obtained
analytically for a large spin system $N = 10^5$ with $\omega_{1}=0.625 \times 10^{-5} $. As for Fig.~\ref{Fig1d5mM0125},
the two curves show different behaviors.% The polarization inversion, for
%the spin-temperature Ansatz, did not emerge
%in Fig.~\ref{Fig1d5mM0125}, but is here observed because of the much larger 
%system sizes.
}
\label{Fig0Analytic}
\end{figure}

Instead, the spin-temperature approach in \eqref{stans} reduces
to assuming $P_z(\omega) = P_z^{\mbox{\tiny ans}}(\omega)  \equiv \tanh(\beta_s 
(\omega - \omega_0)/2)$. Then, 
the parameters $\beta_s$ and $\omega_0$ are obtained from 
Eqs.~\eqref{blochtotal}.
This approach permits us to achieve much larger system sizes. For instance, the 
results at $N = 10^5$ are shown
in Fig.~\ref{Fig0Analytic}, where we plot the stationary polarizations versus 
the resonance frequency $\omega_i = \omega_e + \Delta_i$.
As explained at the end of Sec.~\ref{sec:calc}, this is easily connected to the EPR spectrum.

As expected, the two curves show radically different behaviors. It is 
interesting to observe that 
the spin-temperature curve presents the polarization inversion of Eq.~\eqref{PzST}
but exhibits a shift of the point $\omega_0$ where it crosses the horizontal axis 
with respect to $\omw$. 
This effect, together with the small sizes accessible 
in the numerical simulation of the interacting case, explains why 
the polarization inversion does not clearly appear in Fig.~\ref{Fig1d5mM0125},
where the dipolar interactions are so weak that one would expect
the simplified Borghini model of Eq.~\eqref{PzST} to hold.

\section{Conclusions}

\paragraph*{Comparison with experiments. ---}
In Ref.~\cite{hovav2015electron}, the authors study experimentally the EPR spectrum
under microwave irradiation for pyruvic acid doped with TEMPOL or trityl radicals.
At high temperature ( $ \GtrSim 10$ K),
the shape of the EPR spectrum is well described in terms of hole-burning 
and few-body processes involving protons and $^{13}C$ nuclei (cross effect, solid effect).
At low temperature ($ = 2.7$ K), the EPR spectrum shows a broad depolarization 
which is explained phenomenologically by including an electron spectral diffusion
term. What is the microscopic origin of such a term?

In this work, we computed by exact diagonalization 
the stationary EPR spectrum of $12$ spins in presence of dipolar interaction and standard DNP conditions. 
Under the hypothesis of slow spin-relaxation time $T_1$, which holds only at sufficiently low temperatures,
we show that for the concentrations used in the experiment, the EPR spectrum displays 
a broad depolarization similar to the one observed in \cite{hovav2015electron}. Let us stress that
in our approach, no phenomenological term was added and therefore the presence of electron spectral diffusion
can be directly put in relation with dipolar interactions. 
Moreover, the EPR spectra that we obtained are equivalent to the one that would be measured at equilibrium 
but at an inverse temperature $\beta_s$ and an effective magnetic field $\omega_e - \omega_0$, as shown in Fig.~\ref{Fig:15mM}.
Such a picture breaks down at lower concentrations in agreement with the localization/delocalization picture
introduced in \cite{DeLuca2015} and the EPR spectrum is well described by a non-interacting hole burning. 
We remark that for simplicity, we restricted to a minimal set of parameters, which are nevertheless
enough to describe qualitatively the results of Ref.~\cite{hovav2015electron}; in particular,
we kept the parameters $T_1$ and $T_2$ independent of the
radical concentration, while it is known that they should become longer at low concentrations.
In order to perform an explicit comparison, one would need to take into account
these specific details, including the exact shape of the non-irradiated EPR spectrum.

As discussed in theoretical works \cite{DeLuca2015, de2016thermalization},
the transition separating hole burning and spin-temperature scenarios can be identified 
with a many-body localization, recently identified in the context of 
quantum thermalization~\cite{Basko2006, Pal2010, de2013ergodicity, Nandkishore2015}.
Thanks to the analysis presented here, we suggest that the EPR spectrum 
can be an additional and important tool to detect and investigate this transition in DNP protocols. 
We are confident that this is an intriguing possibility that can pave the way 
for forthcoming experiments aiming at clarifying the microscopic mechanisms underlying DNP.

\paragraph*{Role of glassiness in DNP protocols. ---} 
Our results are also relevant for DNP applications, as our simulations show
important hyperpolarization for radicals arranged in a regular cubic lattice.
This suggests that high level of hyperpolarization can be obtained even 
in presence of a perfect spatial regularity in the radicals positions.
Nevertheless, it is a well established experimental fact, that
a sizable nuclear polarization is in practice only achieved for amorphous 
compounds. How to conciliate these two apparently contradictory observations?
Our conclusion is that the success of hyperpolarization is actually granted by 
a sufficiently homogeneous distribution of the radical in the sample, no matter if regular or random.
However, in polycrystalline samples, it is well known that radical spins
are forced to accumulate at the boundaries between different crystalline grains,
as already suggested by NMR measurements reported in Ref. \cite{amorphous2013}.
Inside these boundary regions, an anomalously high radical concentration is responsible for 
a suppression of hyperpolarization as confirmed 
by experimentally~\cite{johannesson2009dynamic} and 
theoretically~\cite{ColomboSerra2014, DeLuca2015}.

\paragraph*{Acknowledgements. }\hspace{0.1cm}
We thank M. M\"uller and In\`es A. Rodr\'iguez for fruitful discussions. 
We are also glad to thank Y. Hovav and S. Vega for pointing out their 
interesting results
which motivated our work. 
This work is supported by ``Investissements d'Avenir'' LabEx PALM 
(ANR-10-LABX-0039-PALM).

\footnotesize{
\bibliography{dnp} %your .bib file
\bibliographystyle{rsc} %the RSC's .bst file
}

\end{document}